\documentclass[aps,onecolumn,showpacs,amsfonts,eqsecnum,float,prl,preprint]{revtex4} 
\usepackage{epsfig,amsmath}
\usepackage{bm}

\begin{document}

\title{Interference in Bohmian Mechanics with Complex Action}

\author{ Yair Goldfarb and David J. Tannor} \affiliation{Dept. of
Chemical Physics, The Weizmann Institute of Science, Rehovot, 76100
Israel \\ \today}

\begin{abstract}

\noindent In recent years, intensive effort has gone into developing
numerical tools for exact quantum mechanical calculations that are
based on Bohmian mechanics. As part of this effort we have recently
developed as alternative formulation of Bohmian mechanics in which
the quantum action, $S$, is taken to be complex [JCP {\bf125},
231103 (2006)]. In the alternative formulation there is a
significant reduction in the magnitude of the quantum force as
compared with the conventional Bohmian formulation, at the price of
propagating complex trajectories. In this paper we show that
Bohmian mechanics with complex action is able to overcome the main
computational limitation of conventional Bohmian methods
--- the propagation of wavefunctions once nodes set in.
In the vicinity of nodes, the quantum force in conventional Bohmian
formulations exhibits rapid oscillations that pose severe
difficulties for existing numerical schemes. We show that within
complex Bohmian mechanics, multiple complex initial conditions can
lead to the same real final position, allowing for the description
of nodes as a sum of the contribution from two or more crossing
trajectories. The idea is illustrated on the reflection amplitude
from a one-dimensional Eckart barrier. We believe that trajectory
crossing, although in contradiction to the conventional Bohmian
trajectory interpretation, provides an important new tool for
dealing with the nodal problem in Bohmian methods.

\end{abstract}

\maketitle

\renewcommand{\theequation}{\arabic{equation}}



\noindent The challenge of performing quantum mechanical
calculations on systems of many degrees of freedom has been a major
focus of the theoretical chemistry community for almost four
decades. Since classical mechanics can be applied to systems with
tens of thousands of degrees of freedoms it is only natural that
much attention has focused on methods capable of describing quantum
effects but requiring only the propagation of classical
trajectories. One approach that has shown significant progress in
recent years is the use of Bohmian mechanics
(BM)\cite{courtney,corey,trahan,jian,erik,sophya,burgha,ginden,poirier,
babyuk,kend,wyattb}. Originally developed in the 1950's as a causal
formulation of quantum mechanics (QM), in this formulation
trajectories evolve in the presence of the usual Newtonian force
plus an additional quantum force\cite{bohm}. Although the quantum
force is nonlocal, the formulation of BM in terms of trajectories
suggests the possibility of achieving an efficiency that is
compatible with classical trajectory methods. Despite its successes,
BM suffers from several drawbacks that have prevented it from
becoming an effective numerical tool to date, the most significant
of these being the nodal problem --- a numerical instability in
regions where the wavefunction oscillates that eventually leads to a
breakdown of the numerical scheme\cite{wyattb}.

In this paper we address the nodal problem using a recently
developed variation on BM that we call Bohmian mechanics with
complex action (BOMCA) \cite{goldfarb,goldfarb7}, where the phase
$S$, is taken to be complex. Like conventional BM, BOMCA is formally
identical to the exact Schr\"odinger equation; however the quantum
force is significantly smaller and more localized than in
conventional Bohmian mechanics. This comes at the expense of the
trajectories being complex.

In our previous publication\cite{goldfarb} we focused on the
transmitted part of a wavepacket scattered from an Eckart barrier;
we demonstrated tunneling probabilities that were in virtually
perfect agreement with the exact quantum mechanics down to
$10^{-7}$. Here we complete the picture by calculating the reflected
part of the wavefunction. Whereas the transmitted part is nodeless,
the reflected part, for a wide range of energies, has an oscillatory
structure. We show that the oscillatory structure is obtained
automatically in BOMCA simply by including the contributions from
multiple initial conditions that lead to the same final position.


The origin of the nodal problem in BM can be traced back to the
hydrodynamic equations of QM, which constitute the first step in the
derivation of the Bohmian formulation. In conventional BM these
equations take the form:
\begin{eqnarray}
\label{st0}
S_{t}+\frac{S_{x}^{2}}{2m}+V&=&-Q, \\
\label{at0} A_{t}+\frac{1}{m}A_{x}S_{x}+\frac{1}{2m}AS_{xx}&=&0,
\end{eqnarray}
where $V(x,t)$ is the potential of the system,
$m$ is the mass of the particle, $\hbar$ is Planck's constant
divided by $2\pi$ and the subscripts denote partial derivatives.
\begin{equation}
Q\equiv-\frac{\hbar^{2}}{2m}\frac{A_{xx}}{A}
\end{equation}
is referred as
the ``quantum potential".  As
can be seen from the expression for $Q$, the quantum potential
diverges at nodal regions of the wavefunction. Numerically the
difficulty is even more severe
--- well before a node is formed, when the amplitude of the
wavefunction exhibits only nodeless ripples, the quantum
trajectories are highly unstable due to rapid oscillations in the
quantum potential\cite{wyattb}.

Since nodes in quantum mechanics arise from interfering amplitudes,
it is only natural to attempt to solve the nodal problem by applying
the superposition principle --- to decompose the wavefunction into
two nodeless parts and to propagate each part separately using
trajectories. Indeed, two such methods have been developed, the
Counter Propagating Wave Method (CPWM)\cite{poirier} and the
Covering Function Method (CFM)\cite{babyuk}. However, since nodeless
wavefunctions do not stay nodeless for a general potential, these
methods generally require a series of time-dependent decompositions
of the total wavefunction, which is numerically inconvenient and
largely arbitrary.

A somewhat more natural strategy to solve the nodal problem is to
apply the superposition principle directly to the contribution of
the quantum trajectories. The superposition of contributions from
multiple trajectories is a central concept in the semiclassical
literature\cite{miller}; however, in conventional BM, the crossing
of trajectories in configuration space is strictly prohibited.
Indeed, the "no-crossing" rule plays a central role in conventional
BM: if trajectories could cross it would undermine the Bohmian
interpretation of QM, in which the quantum trajectories are
candidates for the actual trajectory on which a particle propagates
in space. Since the BOMCA formulation yields generally an
approximation to QM, trajectories are allowed to cross. In this
paper we demonstrate how combining the contributions of crossing
trajectories yields accurate interference patters.
%

The starting point of the BOMCA formulation\cite{goldfarb} is the
insertion of the ansatz\cite{pauli,kurt,leacock,david}
$\psi(x,t)=\exp{\left[\frac{i}{\hbar}S(x,t)\right]}$ in the
time-dependent Schr\"odinger equation, where we allow the phase to
be \textit{complex}. This yields a \textit{single} quantum complex
Hamilton-Jacobi equation\cite{pauli,kurt,leacock,david}
\begin{equation}
\label{st1}
 S_{t}+\frac{1}{2m}S^{2}_{x}+V=\frac{i\hbar}{2m}S_{xx},
\end{equation}
In the spirit of conventional BM our aim is to solve eq.(\ref{st1})
in the Lagrangian approach, that is along quantum trajectories. A
quantum trajectory is defined by
\begin{equation}
\label{dxdt1}
 \frac{dx}{dt}=v(x,t); \ \ \ v(x,t)\equiv\frac{1}{m}S_{x}(x,t).
\end{equation}
Due to the definition of $x$ as time-dependent in eq.(\ref{dxdt1}),
we write the solutions of this equation as $x(t;x_{0})$ where
$x_{0}$ is the starting point of the trajectory. Unlike conventional
BM, the complex value of $S$ yields quantum trajectories
$x(t;x_{0})$ that evolve in the complex plane. A Newtonian-like
equation of motion for $v(x,t)$ is obtained by taking a spatial
derivative of eq.(\ref{st1}) and applying eq.(\ref{dxdt1}); after a
short manipulation we obtain
\begin{equation}
\label{dv}
\frac{dv[x(t;x_{0}),t]}{dt}=\underbrace{-\frac{V_{x}}{m}}_{F_{c}/m}+\underbrace{\frac{i\hbar}{2m}v_{xx}}_{F_{q}/m},
\end{equation}
where we identify $F_{c}$, $F_{q}$ as the classical and the quantum
force respectively. The presence of $v_{xx}$ in the quantum force
term prevents the first equation in (\ref{dxdt1}) and eq.(\ref{dv})
from being a closed set.

We tackle the problem of calculating $v_{xx}$ along a trajectory by
taking iterated spatial partial derivatives of eq.(\ref{dv}). After
a short manipulation the result can be written as
\begin{equation}
\label{set_dvdt}
 \frac{dv^{(n)}}{dt}=-\frac{V^{(n+1)}}{m}+\frac{i\hbar}{2m}v^{(n+2)}-\tilde{g}_{n}
  ; \ \ n=0,...\infty,
 \end{equation}
where $\tilde{g}_{0}=0$ and
$\tilde{g}_{n}=\sum_{j=1}^{n}\binom{n}{j}v^{(j)}v^{(n-j+1)}$ for
$n\geq 1$. Here $v^{(n)}$ denotes the $n^{th}$ spatial derivative of
$v$. A similar procedure was used in Refs.\cite{jian,trahan,corey}
in conventional BM. The set of eqs.(\ref{set_dvdt}) and the first
equation in (\ref{dxdt1}) are now an infinite but closed set that
describes a \textit{fully local} complex quantum trajectory. We may
obtain a numerical approximation by truncating the infinite set at
some $n=N$, thus replacing eq.(\ref{dv}) with a system of $N+1$
coupled ODEs. Since each equation of motion for $v^{(n)}$ in
(\ref{set_dvdt}) depends on the subsequent $v^{(n+2)}$, the
truncation is done by setting $v^{(N+1)}=v^{(N+2)}=0$. The initial
conditions for the $v^{(n)}$'s are given by
\begin{equation}
\label{initial}
 v^{(n)}(0;x_{0})=\left.\frac{1}{m}\frac{\partial^{n}S_{x}(x,0)}{\partial
 x^{n}}\right|_{x=x_{0}}=\frac{\partial^{n}}{\partial
 x^{n}}\left.
 \left[-i\hbar\frac{\psi_{x}(x,0)}{\psi(x,0)}\right]\right|_{x=x_{0}},
\end{equation}
where we have applied the definition from (\ref{dxdt1}) together with
$S=-i\hbar\ln{\psi}$, the latter following from the original ansatz.
$x_{0}$ is the initial position of an arbitrary single trajectory,
which is generally complex. The equation of motion for the action
along a trajectory is similar to its classical counterpart, with the
addition of the quantum potential:
\begin{eqnarray}
\label{dsdt1}
\frac{dS[x(t;x_{0}),t]}{dt}=S_{t}+vS_{x}=\frac{1}{2}mv^{2}-V+\frac{i\hbar}{2}v_{x}.
\end{eqnarray}
Solving for $S$ and inserting the result into the original ansatz yields the wavefunction
$\psi[x(t;x_{0}),t]=\exp{\{\frac{i}{\hbar} S[x(t;x_{0}),t]}\}$ at
position $x(t;x_{0})$ in the complex plane. In Ref. \cite{goldfarb},
we discussed the practical issue of finding trajectories that end up
on the real axis $x(t_{f};x_{0})\in\mathbb{R}$ at $t_{f}$. Once
these trajectories are found, the wavefunction on the real axis
$\psi[x(t_{f};x_{0}),t_{f}]$ is reconstructed.
%

As a numerical example, consider the one-dimensional scattering of
an initial Gaussian wavepacket
$\psi(x,0)=(2\alpha/\pi)^{1/4}e^{\left[-\alpha(x-x_{c})^{2}+\frac{i}{\hbar}p_{c}(x-x_{c})\right]}$
from an Eckart potential $V(x)=D/ \cosh^{2}(\beta x)$. We take
$x_{c}=-0.7$, $\alpha=30\pi$, $D=40$, $\beta=4.32$ and $m=30$ (all
units are atomic units). The system is the same as in our previous
publication\cite{goldfarb}. The average translational energy of the
initial Gaussian is taken as $E=p_{c}^{2}/m=10<D$. We focus on
trajectories that end up at a final time $t_{f}=0.995$ with
$x(t_{f})<0$, and thus contribute to the reflected part of the
wavefunction. $t_{f}$ is chosen as sufficiently long for the
wavepacket to scatter from the barrier and interference effects to
appear. First we focus on the $N=1$ BOMCA approximation, for which
the equations of motion are:
\begin{equation}
\label{set1}
\frac{dx}{dt}=v, \ \ \ \ \frac{dv}{dt}=-\frac{V_{x}}{m}, \ \ \ \
\frac{dv_{x}}{dt}=-\frac{V_{xx}}{m}-v_{x}^{2},
\end{equation}
with the auxiliary eq.(\ref{dsdt1}). Note that the $N=1$
trajectories obey Newton's equations of motion. As indicated in Ref.
\cite{goldfarb}, eqs.(\ref{set1}) are closely related to Generalized
Gaussian Wavepacket Dynamics (GGWPD)\cite{huber2}, which also uses
complex trajectories.  As a result, many of the insights in Ref.
\cite{huber2} concerning multiple root trajectories can be carried
over to the case of complex Bohmian mechanics, although we have
found that the structure and the number of the root branches
generally depend on the value of $N$, the order of truncation.
\begin{figure}[h]
\begin{center}
\epsfxsize=8.5 cm \epsfbox{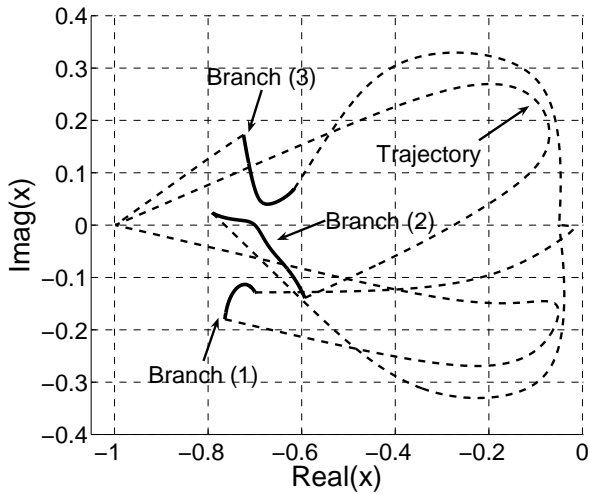} 
\end{center}
\caption{\label{fig_tra_1} Scattering of a Gaussian from an Eckart
potential barrier within the $N$=1 BOMCA approximation (the
parameters are given in the text). Three branches and six sample
complex trajectories are depicted. The branches are the locus of
initial positions of trajectories that end at time $t_{f}=0.995$ at
real $x_{f}$, $-1<x_{f}<-0.05$, and thus contribute to the reflected
wavefunction. The figure shows two sample trajectories emerging from
each branch, one that ends at $x_{f}=-1$ and one that ends at
$x_{f}=-0.05$.}
\end{figure}

In fig.\ref{fig_tra_1} we plot three branches that contribute to the
reflected wavefunction. A branch is defined as the locus of
\textit{initial} positions of trajectories that end at time $t_{f}$
at real $x_{f}$ with $x_{f}$ corresponding to a reflected segment of
the wavefunction, $x_{f}\in[-1,-0.05]$. Two sample trajectories are
depicted emerging from each branch, one that ends at position
$x_{f}=-1$ and one that ends at $x_{f}=-0.05$. Thus, to each final position
(for this segment of the reflected wavefunction) there correspond three
initial positions --- one originating from each of the three branches
--- and the wavefunction should therefore include contributions from all
three branches. As in Ref. \cite{huber2}, at short times only one
branch contributes to the final wavefunction. We refer to this
branch as the ``real branch" since it incorporates a trajectory that
stays on the real axis at all times (the trajectory that initiates
from $x_{0}=x_{c}$). At longer times, secondary branches begin to
make a significate contribution to the final wavefunction. There is
apparently no fundamental limitation on the number of the secondary
branches, although in the section of the complex plane depicted in
fig.\ref{fig_tra_1} no other branches were found. In
fig.\ref{fig_tra_1}, branch (2) is the real branch while branches
(1) and (3) are secondary branches.
\begin{figure}[h]
\begin{center}
\epsfxsize=8.5 cm \epsfbox{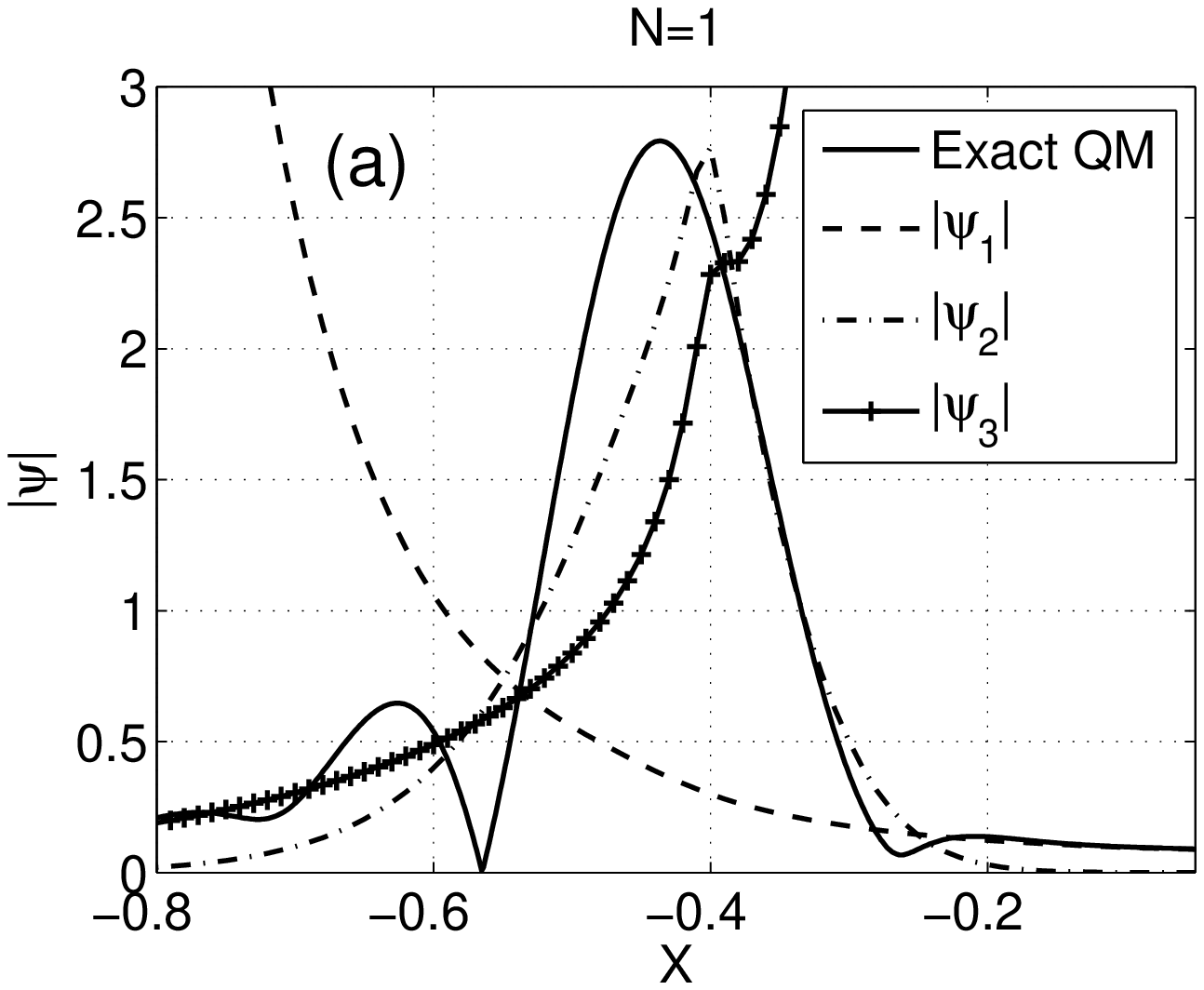} 
\epsfxsize=8.5 cm \epsfbox{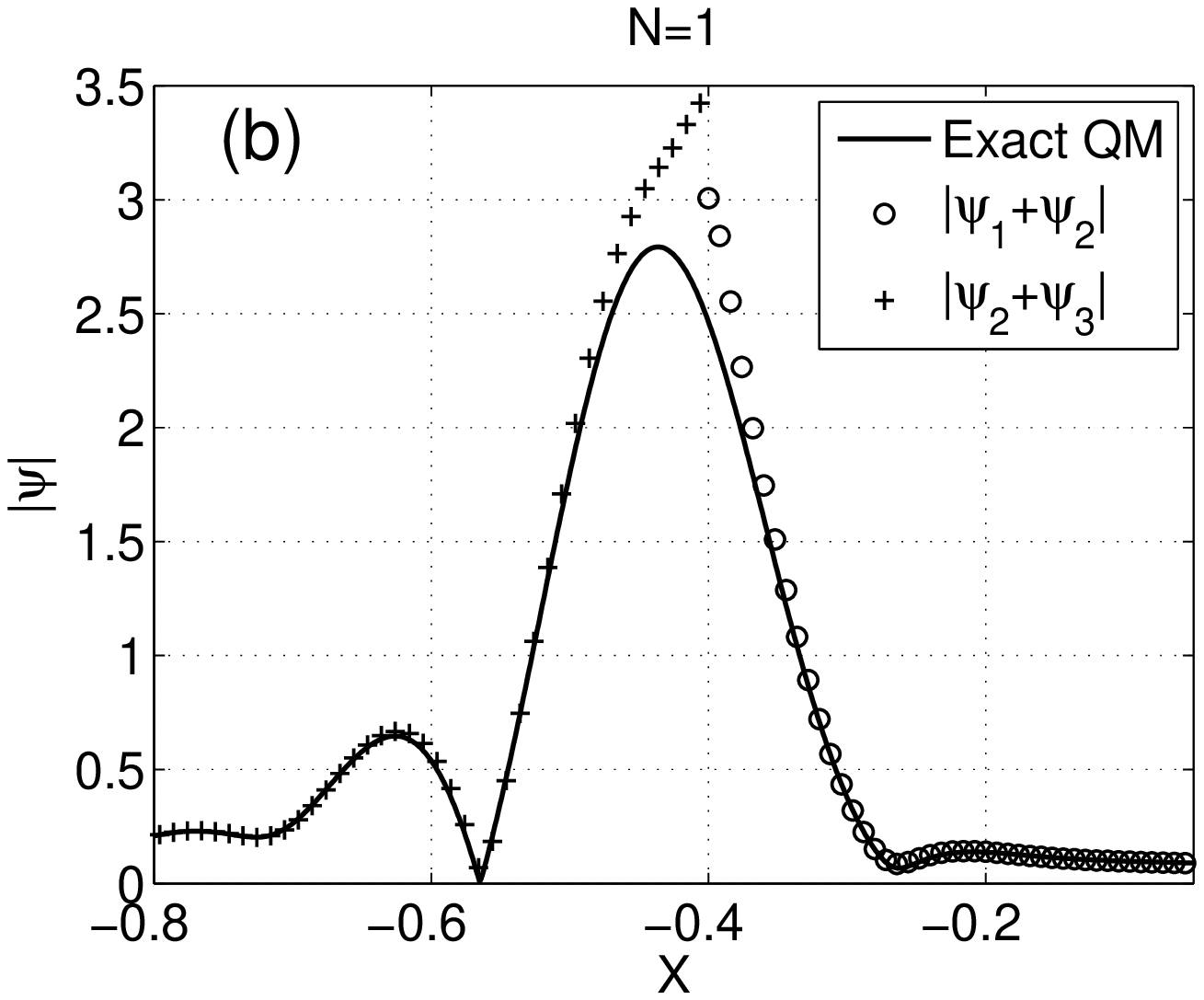} 
\end{center}
\caption{\label{fig_psi_1} (a) $N=1$ BOMCA approximation
corresponding to each of the three branches $j=1,2,3$ (see
fig.\ref{fig_tra_1}),
$|\psi_{j}(x,t_{f})|=|\exp[iS_{j}(x,t_f)/\hbar]|$. In the region of
the transmitted wavefunction (starting from $x\approx-0.2$), a good
approximation to the exact result is obtained by just a single
branch (branch (1)). (b) The result of adding contributions from
pairs of branches $|\psi|=|\psi_{j}+\psi_{i}|, \ \ i\neq j$.
Different choices of $i$ and $j$ are useful in different regions of
the wavefunction.}
\end{figure}

In fig.\ref{fig_psi_1}(a) we plot
$|\psi_{j}(x,t_{f})|=|\exp[iS_{j}(x,t_f)/\hbar]|,\ j=1,2,3$, where
$j$ corresponds to each of the three branches.
In fig.\ref{fig_psi_1}(b) we present the result of
adding pairs of branches, $|\psi|=|\psi_{j}+\psi_{i}|, \ \ i\neq j$.
We see that to a good approximation the rippled parts of the
wavefunction are described by a simple superposition with a proper
choice of $i$ and $j$.

Note that as we approach the region of the transmitted part of the
wavefunction ($x \gtrsim -0.2$) a \textit{single} branch (branch
(1)) is enough to provide a good approximation.  In fact, this is
the branch that is responsible for the transmitted wavefunction
($x_{f}>0$, not shown), and hence we will call this the "transmitted
branch". At energies on the order of magnitude of the barrier
height, the ``transmitted branch" is the real branch, but at lower
energies (and longer time scales) there is a crossover and the
transmitted branch is one of the secondary branches. The existence
of a single transmitted branch coincides well with the nodeless
character of the transmitted part of the wavefunction.  By the same
token, in Ref. \cite{goldfarb} we showed that the reflected part of
the wavefunction can be well approximated by a single branch if it
has no ripples or nodes.

The issue of which branches should be included in the sum and when,
is obviously of central importance to the method; at present, we do
not have a rigorous justification for the neglect of certain
branches.  A partial discussion is given in Ref. \cite{huber2} in
the context of GGWPD, but the BOMCA formulation, being more general,
requires a more comprehensive discussion, which we leave to a future
publication.

\begin{figure}[h]
\begin{center}
\epsfxsize=8.5 cm \epsfbox{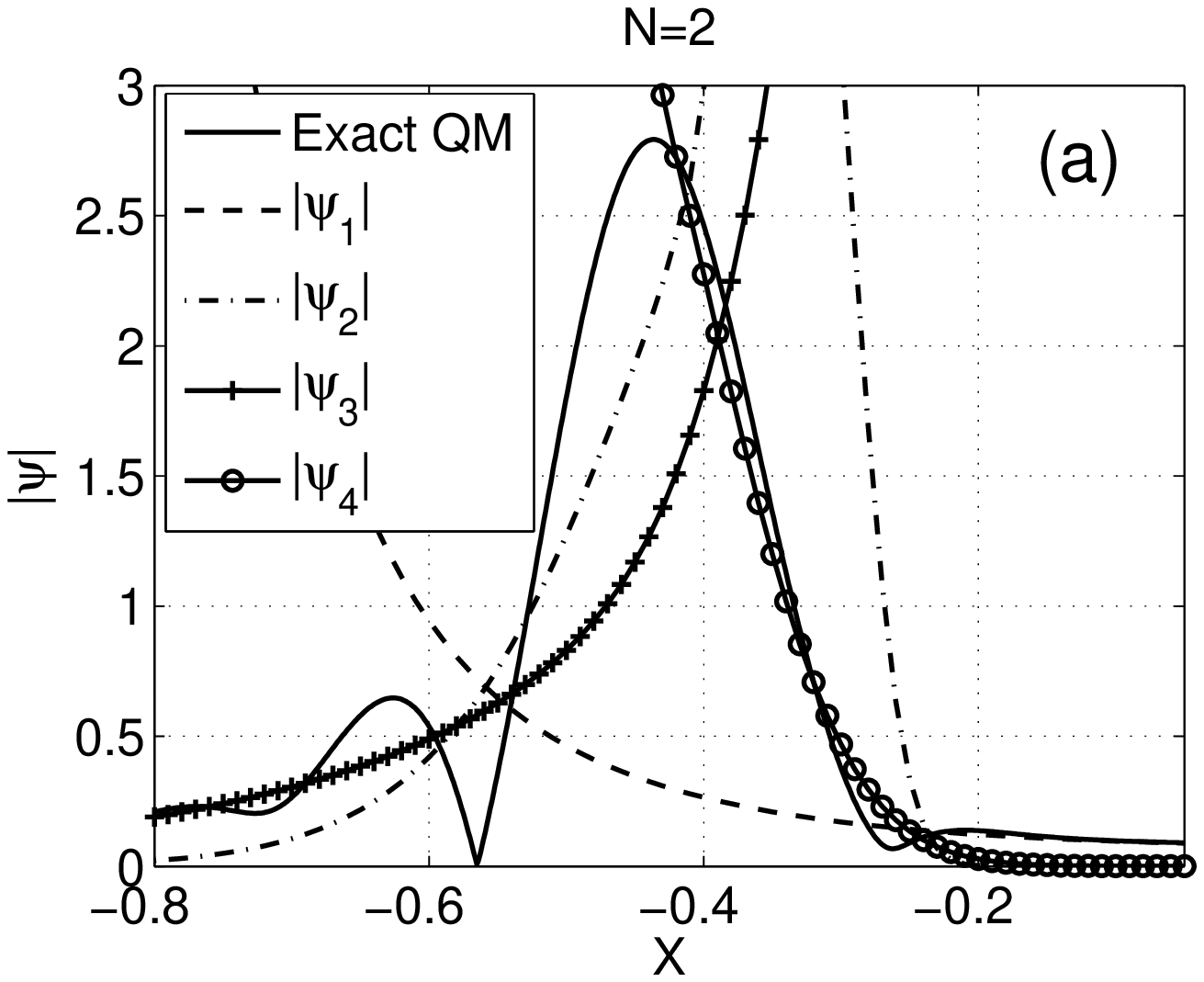} 
\epsfxsize=8.5 cm \epsfbox{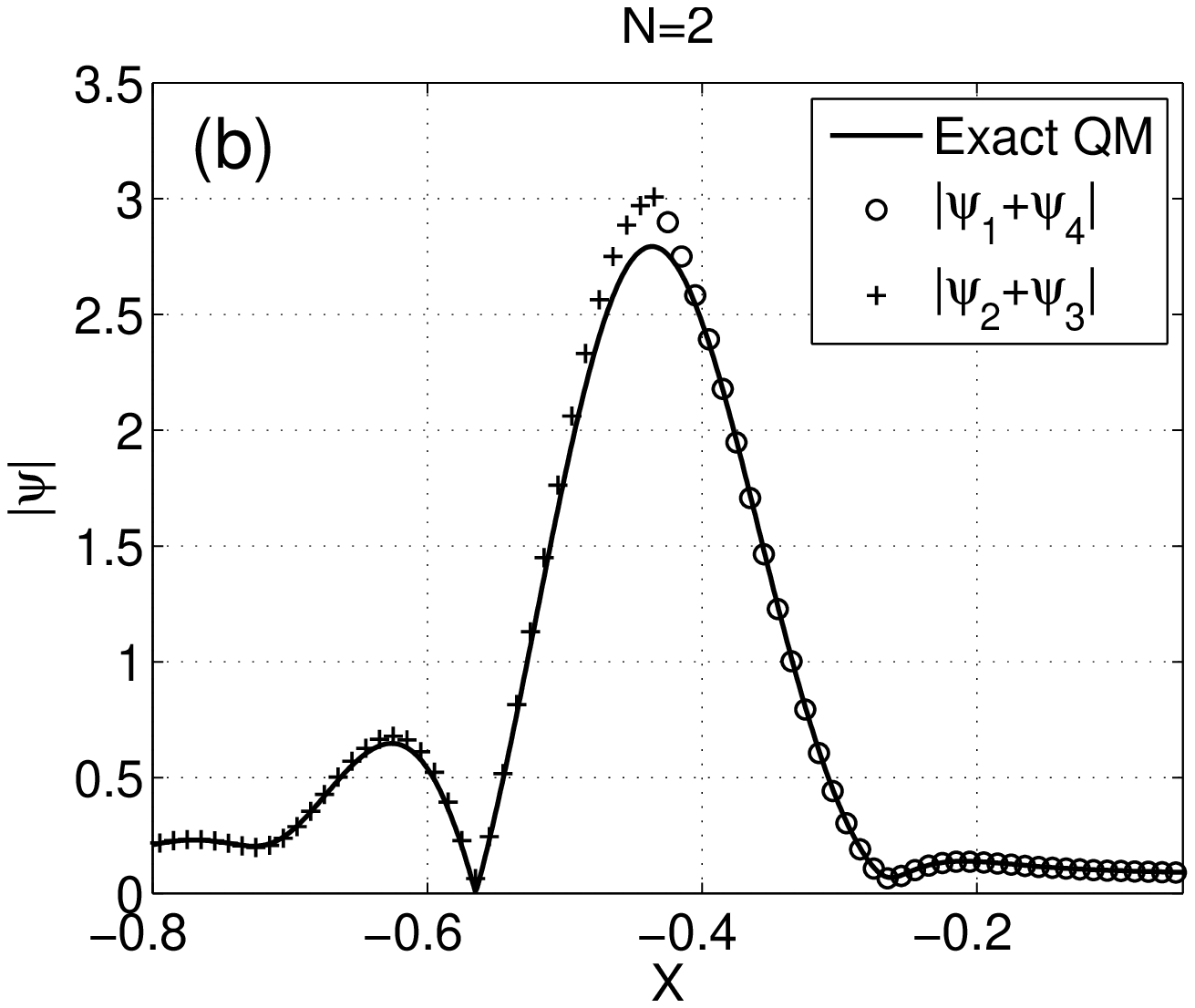} 
\end{center}
\caption{\label{fig_psi_n2_1} (a) $N=2$ BOMCA approximation
corresponding to four branches
$|\psi_{j}(x,t_{f})|=|\exp[iS_{j}(x,t_f)/\hbar]|$, $j=1,...,4$. (b)
The result of adding contributions from pairs of branches
$|\psi|=|\psi_{j}+\psi_{i}|, \ \ i\neq j$. Except for a slight
decrease in accuracy left to the maximum, the results for $N=2$
where a quantum force term is added are better than for $N=1$
(complex classical trajectories).}
\end{figure}

We now turn to the next order of approximation, $N=2$. The equations
of motion are based on eqs.(\ref{set1}) with two changes. The second
equation in (\ref{set1}) is replaced with eq.(\ref{dv}), in which
there is a quantum force. Knowledge of the quantum force requires an
equation of motion for $v_{xx}$. We derive such an equation by
inserting $n=2$ in eqs.(\ref{set_dvdt}) and taking $v^{(4)}=0$,
leading to
\begin{equation}
\frac{dv_{xx}}{dt}=-\frac{V_{xxx}}{m}-3v_{x}v_{xx}.
\end{equation}
At the level of $N=2$, four branches contribute to the reflected
wavefunction at $t_{f}$. Fig.\ref{fig_psi_n2_1}(a) depicts the
contribution of the four branches, $|\psi_{j}|$, $j=1,..,4$. The
result of adding pairs of branches $|\psi|=|\psi_{j}+\psi_{i}|, \ \
i\neq j$ is given in fig.\ref{fig_psi_n2_1}(b). There is a
significant increase in accuracy relative to $N=1$ in the vicinity
of the maximum and to its right, although there is a slight decrease
in accuracy relative to $N=1$ to the left of the maximum.

In conclusion, we have demonstrated that BOMCA accounts for
interference and nodal structures of wavefunctions in a simple and
natural way. In spite of the conceptual difficulties that crossing
trajectories may pose for BM as an interpretational tool of QM, this
notion introduces a powerful numerical tool and might even enrich
the orthodox interpretation of the Bohmian formulation. The present
results, combined with the simple and accurate results obtained for
tunneling in Ref. \cite{goldfarb}, demonstrates great promise for
BOMCA as a versatile alternative to current semiclassical methods.
As mentioned above, several issues require a more comprehensive
understanding. First, what are the convergence properties of the
method as higher order approximations are taken to the quantum
force, that is increasing the value of $N$? Can rigorous rules be
derived for the summation of the different branches? What is the
relation between the exact phase that diverges at a node and the
approximate BOMCA formulations that can account for nodes via a
bipolar or multipolar expansion? We intend to address all these
questions in future work. This work was supported by the Israel
Science Foundation $(576/04)$.




%


\begin{thebibliography}{99}
\bibliographystyle{phaip}
%
%
\bibitem{courtney} C. L. Lopreore, R. E. Wyatt, Phys. Rev. Lett. {\bf82}, 5190 (1999).
\bibitem{corey} C. J. Trahan, K. Hughes, R. E. Wyatt, J. Chem. Phys. {\bf118}, 9911 (2003).
\bibitem{trahan} C. J. Trahan, R. E. Wyatt, B. Poirier, J. Chem. Phys. {\bf122}, 164104 (2005)
\bibitem{jian} J. Liu, N. Makri, J. Phys. Chem. A {\bf108}, 5408 (2004).
\bibitem{erik} E. R. Bittner, R. E. Wyatt, J. Chem. Phys. {\bf113}, 8888 (2000).
\bibitem{sophya} S. Garashchuk, V. A. Rassolov, Chem. Phys. Lett. {\bf364}, 562 (2002).
\bibitem{burgha} I. Burghardt, L. S. Cederbaum, J. Chem. Phys. {\bf115}, 10303 (2002).
\bibitem{ginden} E. Gindensperger, C. Meier and J. A. Beswick, J. Chem. Phys. {\bf113}, 9369 (2000).
%
\bibitem{poirier} B. Poirier, J. Chem. Phys. {\bf121}, 4501 (2004).
\bibitem{babyuk} D. Babyuk, R. E. Wyatt, J. Chem. Phys. {\bf121}, 9230 (2004).
\bibitem{kend} B. K. Kendrick, J. Chem. Phys. {\bf119}, 5805 (2003).
\bibitem{wyattb} R. E. Wyatt, \textit{Quantum Dynamics with
Trajectories: Introduction to Quantum Hydrodynamics} (Springer, New
York, 2005).
%
%
\bibitem{bohm} D. Bohm, Phys. Rev. {\bf85}, 166 (1952); D. Bohm, Phys. Rev. {\bf85}, 180 (1952).
%
\bibitem{goldfarb} Y. Goldfarb, I. Degani, D.J. Tannor, J. Chem. Phys. {\bf125}, 231103 (2006).
%
\bibitem{goldfarb7} Y. Goldfarb, J. Schiff, D. J. Tannor, J. Phys. Chem (submitted).
%
%
\bibitem{miller} W. H, Miller, \textit{Advances in Chemical Physics}
{\bf{25}}, 69 (1974).
%
\bibitem{pauli} W. Pauli, \textit{Die allgemeine Prinzipien der
Wellenmechanik}, Handbuch der Physik, Vol. 24, Part 1, 2nd ed.,
Springer-Verlag, Berlin, 1933.
\bibitem{kurt} K. Gottfried \textit{Quantum Mechanics, Volume
I: Foundations} (W. A. Benjamin, New-York, 1966).
%
\bibitem{leacock} R. A. Leacock, M. J. Padgett, Phys. Rev. D
{\bf28}, 2491 (1983).
%
\bibitem{david} D. J. Tannor, \textit{Introduction to Quantum Mechanics: A Time Dependent
Perspective} (University Science Press, Sausalito, 2006).
%
\bibitem{huber2} D. Huber, E. J. Heller, R. G. Littlejohn, J. Chem. Phys. {\bf89}, 2003 (1988).
%
%
%
%
%
%
%
%



%
\end{thebibliography}
\end{document}